\documentclass[pdflatex,sn-mathphys]{sn-jnl}
\usepackage{appendix}
\usepackage{tabularx}
\usepackage{booktabs}
\usepackage{hyperref}
\usepackage{subcaption}

\usepackage{float} 
\usepackage{soul,color}
\DeclareUnicodeCharacter{0301}{}
\usepackage[flushleft]{threeparttable}

\jyear{2022}%

\theoremstyle{thmstyleone}%

\begin{document}

\title[The Framework For The Discipline Of Software Engineering]{The Framework For The Discipline Of Software Engineering in Connection to Information Technology Discipline}

\author*[1]{\fnm{Jones} \sur{Yeboah}}\email{yeboahjs@mail.uc.edu}
\author[2]{\fnm{Feifei} \sur{Pang}}\email{pangfi@mail.uc.edu}
\author[3]{\fnm{Hari Priya} \sur{Ponnakanti}}\email{ponnakha@mail.uc.edu}

\affil*[1]{\orgdiv{School of Information Technology}, \orgname{University of Cincinnati}, \orgaddress{\street{2600 Clifton Avenue}, \city{Cincinnati}, \postcode{45221}, \state{Ohio}, \country{USA}}}

\begin{huge}
\textbf{Springer Copyright Notice}
\end{huge}

\vspace{5mm} 

\begin{large}
Copyright(c) 2022
\end{large}
\vspace{5mm} 
\begin{large}

This work is subject to copyright. All rights are reserved by the Publisher, whether the whole or part of the material is concerned, specifically the rights of translation, reprinting, reuse of illustrations, recitation,broadcasting, reproduction on microfilms or in any other physical way, and transmission or information storage and retrieval, electronic adaptation, computer software,or by similar or dissimilar methodology now known or hereafter developed.\\

\textbf{Accepted to be published in: }The 20th Int'l Conf on Software Engineering Research and Practice  (SERP'22: July 25-28, 2022, USA)

\end{large}

\vspace{5mm} 

\abstract{
This paper represents preliminary work in identifying the foundation for the discipline of Software Engineering and discovering the links between the domains of Software Engineering and Information Technology (IT). Our research utilized IEEE Transactions on Software Engineering (IEEE-TSE), ACM Transactions on Software Engineering and Methodology (ACM-TOSEM), Automated Software Engineering (ASE), the International Conference on Software Engineering (ICSE), and other related journal publication in the software engineering domain to address our research questions. We explored existing frameworks and described the need for software engineering as an academic discipline. We went further to clarify the distinction difference between Software Engineering and Computer Science. Through this efforts we contribute to an understanding of how evidence from IT research can be used to improve Software Engineering as a discipline. 
}

\keywords{Software Engineering, Information Technology, Framework, Academic Discipline}

\maketitle

\section{Introduction}
Software engineering is the systematic application of engineering approaches to the development of software \cite{b1, b35}. It has two parts: software and engineering. Software is a collection of instructions or codes, documents, and triggers that do a specific job to fill a specific requirement \cite{b1}. They also consists of programs, databases, and documentation, that adds value to the hardware components of a computer, which plays a significant role in all aspects of today’s human daily life \cite{b3}.  Engineering on the other hand is the use of proven systematic principles to design and build machines, structures, and other items, using the best practices, and methods \cite{b34}.


SE pioneer Barry W. Boehm points out that it also includes associated documentation needed for this process \cite{b4}. Practically, SE has been a more engineering activity of considerable importance in the research, creative design, building, test, and management of all software-intensive systems for more than five decades \cite{b2}. The term software engineering was an embodiment of services offered by companies in the 1960s \cite{b10}, and in 1965 it was first referenced \cite{b16}. It was formerly used in 1966 communication of ACM by Oettinger to make the distinction between computer science and the building of software-intensive systems \cite{b11}. SE was also linked to the NATO conference title by Professor Fritz Bauer in the first conference of software engineering in 1968 \cite{b12}. In the 1970s, significant conferences on software engineering were organized, and in the 1980s, software engineering courses were incorporated into specific computer science degree programs \cite{b26}. Oftentimes people confuse computer science (CS) with software engineering and use the two terms interchangeably due to both CS and SE belonging to the computing domain, and both are important for the development of software systems. However, we should not confuse CS with SE as the focus of CS is different from SE’s focus. Broadly, SE focuses on rigorous techniques for designing and producing things that consistently perform what they are meant to do (without affecting the complete network or system). In contrast, CS focuses on creating new knowledge of computers \cite{b23}.

Our research utilizes peer-reviewed papers in computing, engineering, software development, and information technology to address the following research questions:
\begin{itemize}
  \item \textbf{RQ1:} \textit{What is the framework for the discipline of Software Engineering?}
  \item \textbf{RQ2:} \textit{What is the impact of SE in technology development, and how is it related to the CS and IT discipline?}
\end{itemize}

\section{Background Literature}
In this section we reviewed articles from the SE, CS, and IT disciplines to address our research questions by surveying relevant literature from Springer, ACM Transactions on Software Engineering and Methodology, IEEE transactions, ScienceDirect, Clarivate, and other peer-reviewed journals using the following targeted keywords, Software Engineering Theory and Practice, Intersection of Technology and SE, Framework for SE, Software Engineering Academic Discipline, etc.
The increase in the production of personal computers led to more companies producing commercial software for individual users and corporate organizations. SE has been one of the most profitable industries with rapid growth between 1960-1970 \cite{b25}. Many suggest that the term “software engineering” was coined by the North Atlantic Treaty Organization (NATO) Science Committee conference in 1968 to describe a workshop of software production state and prospects assessment as a statement of aspirations, and achieved its popularity in the 1970s \cite{b7}. Over the years, software organizations begin to fragment into other sub-industries such as business applications, project management applications, personal applications, systems software, embedded software, middleware, scientific and mathematical software, communications software, manufacturing software, database software, and software for games and entertainment \cite{b10}. The concept of time-sharing emerged in 1970 to lower the cost of computing and software applications. Most businesses have similar software needs and data processing requirements.\\\\
Software engineering began in the 1960s and it has evolved into a profession with the aim of producing quality software. Quality can be based on the maintenance of software products with other factors such as usability, readability, security, cost, robustness, scalability, and meet customer satisfaction \cite{b30}. In 1969, there was a major shift that IBM unbundled many programs to separate software from hardware \cite{b31}. Other software products such as SAS (Statistical Analysis Software) were created in the year 1960 by the SAS Institute, starting January 1960, SAS was used for data management, business intelligence, predictive analysis, descriptive and prescriptive analysis, etc. Since then, many new statistical procedures and components have been introduced in the field of software engineering. Before 1960, women were mostly assigned the role of software developers, and legends such as Grace Hooper and Margaret Hamilton were pioneers in the programming world. From 1965 to 1985, there was a major crisis that showed many challenges faced in the software engineering field, most issues were centered around budget and schedules as well as property damage. In terms of crisis, we view software based on its productivity but as time evolves the emphasis was on quality as a lot of organizations were having difficulty filling software developer positions \cite{b29}. In early 1980, software engineering began to be on par with computer science and other traditional engineering disciplines. From 1985 to 1989, most researchers and organizations embarked on building software tools to solve the software crisis \cite{b28}. Currently, software engineering has evolved into a broader discipline with tools and environment making it easier to build software with ease \cite{b28}.

\subsection{Software Engineering as an Academic Discipline}
Software engineering as an academic discipline began to evolve in 1984 when the first software engineering program was hosted in Carnegie Mellon University in Pittsburgh, Pennsylvania, United States by government-funded research in 1984 \cite{b13}. Imperial College London hosted the first bachelor’s software engineering program in 1987 at the department of computing \cite{b15} and Rochester Institute of Technology (RIT) was the first school in the United States to offer a software engineering degree program in 1996 \cite{b14}.\\
In 2009, Jacobson and the team identified the huge gap between academic research and the industrial application of software engineering, one of the challenges identified was the lack of a theoretical basis to compare, evaluate and validate experiments. They initiated the Software Engineering and Method (SEMAT) with the aim of founding SE as an academic discipline to establish a general and widely accepted theoretical framework \cite{b8}\\\\
Regarding the confusion aforementioned between CS and SE, CS focused more on the theory of science and digital systems with no emphasis on skills set for designing and creating software that meets users' needs/requirements \cite{b17}. CS is different from SE mainly because SE is perceived equally as an art and science of practically balancing costs, scheduling, complexity, identifying functionality, evaluating performance, as well as legal and ethical forces in software applications \cite{b18}. The software engineering curricula in the RIT hinged on seven main themes namely: Design solutions to customer problems, Software development process, Evolution and Maintenance of software system, Complexity Management, Software standards, Team-based development, and Professionalism in engineering practices \cite{b14}.\\\\
Foster defined SE as a discipline that is concerned with the research, evaluation, analysis, design, construction, implementation, and management of software systems \cite{b3}. The re-engineering of existing software systems with a perspective to improving their role, function, and performance is also a crucial part included in the discipline. Certain fundamental sciences are reflected in the formation and development of SE discipline, including but not limited to principles of mathematics, information science, and computer science \cite{b6}. Specifically, Lavrischeva gave several examples of how SE disciplines could be constructed based on different approaches and scientific foundations of the corresponding fundamental sciences. For instance, SE can be classified as a scientific discipline based on classical sciences, such as algorithm theory and mathematical logic, and the theory of programming\cite{b6}.\\\\
However, although changes are taking place rapidly and significantly in computing and software development during the recent decades, certain fundamental concepts that constitute the SE discipline have remained the same \cite{b23}. Wasserman points out eight key concepts of abstraction; analysis and design methods and notations; user interface prototyping; modularity and architecture; software life cycle and process; reuse; metrics; and automated support. Taken together, they sum up and comprise an applicable foundation for SE as a discipline\cite{b23}.

\subsection{The Framework of Software Engineering}
The framework of software engineering keeps evolving as the timelines in computing move from the pre- mechanical era to the present dispensation where computer systems need written quality software and are also concerned about how to create it. The software engineering framework has external stakeholders, people/team, technology to use, software system and standardize principles/way of working (SDLC and methodological approach) as shown in Figure 1 \cite{b8}. According to the researchers, building quality software's were the core of software engineering research, which includes the external stakeholders that represent organizations or users who are in need of the software solutions to perform their work, the team who are responsible for the development of the software solution, the software company understanding the requirements of the user/external stakeholders and building solutions that meet their needs, also ensuring that lay down software engineering design principles, SDLC and standardized approach for software development life cycle are followed to deliver quality software's that meets industry standards. The effectiveness of these components is what makes software engineering evolving with innovative ideas and techniques to maximize the quality of the software to create.

\begin{figure}[h!]
     \centering
     \begin{subfigure}[b]{0.4\textwidth}
         \centering
         \includegraphics[width=\textwidth]{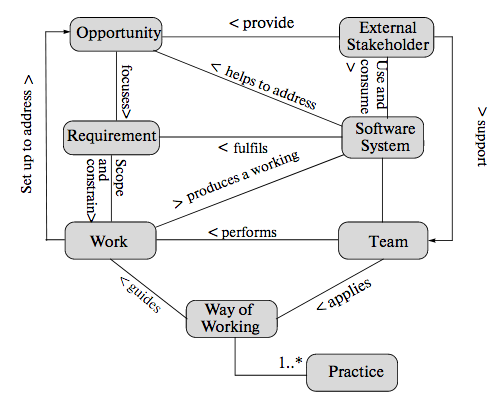}
          \caption{Jacobson et al. \cite{b8}}

         \label{fig: Framework for software engineering}
     \end{subfigure}
     \begin{subfigure}[b]{0.4\textwidth}
         \centering
         \includegraphics[width=\textwidth]{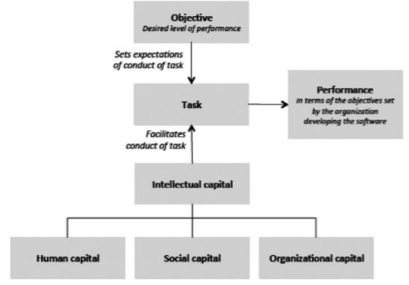}
         \caption{ Wohlin et al. \cite{b22}}
    
         \label{fig: Framework for software engineering 2}
     \end{subfigure}
        \caption{The Framework for software engineering}
        \label{fig:framework}
\end{figure}

The lack of a theoretical basis and a standard benchmark to compare experimental evaluation and validation geared professionals in developing a framework that provides common grounds for researchers in comparing methods and making better decisions \cite{b8}. SE involves understanding the nature of problems and developing techniques for analyzing the problem domain. The Software Engineering framework hinges on four building blocks which are the Methods, Tools (Instruments for accomplishing tasks and enhancing quality), Procedure (Combination of tools and techniques for creating products), and finally Paradigm a unique approach for building software. \cite{b19}.\\\\
Various researchers have proposed different frameworks for SE as a scientific discipline, there is no generally accepted framework for the SE field \cite{b25}. Jacobson et al., proposed a framework that consists of Opportunity, Requirements, Work, Stakeholders and Software System as shown in Figure 1~\cite{b8}. The research of Wohlin et al., also provided a new framework generated from empirical observations of industry practice with three Intellectual Capital Components (ICCs) of Human (Technical Skills and knowledge), Social (Creativity and Network ties), and Organizational (Software, Documentation and Organization culture) Capital as shown in Figure 2. The framework interacts between what set of goals to be achieved (Task), how it should be achieved (Objective), and the combination of ICCs. The task to be achieved remains in the center because it is determined based on the given objectives and the intellectual capital available \cite{b22}.

\section{Relationship Between SE and IT}

Throughout this paper, we have reviewed the history and definition of Software Engineering, which is primarily concerned with the application of engineering principles to the development of software. Information Technology, on the other hand, is largely focused on how computers are used to create, process, store, retrieve, and exchange information. Therefore, the framework of IT according to Said et. al., in 2021 \cite{b33} 
constituted of four connected components: People, Information, Technology, and the Solutions or needs that connect them. The aim of IT discipline in practice is the selection, creation, integration, application, security, and administration of solutions that use technology to empower people through information within organizational or societal contexts, while in research it is the investigation, discovery, and dissemination of needs that connect People, Information, and Technology \cite{b33}.\\\\ 
The similarities and differences are obvious. The common denominator between the two disciplines is the inclusion of people, data/information, and tools/technology. The characteristics of meeting human needs by IT can be combined with software engineering practices to deliver systems that are human centered.  

\section{Discussion}
In our study, we identified that the field of SE involves using knowledge of computers in solving critical human problems, and understanding the nature of the problem is the first significant step in using tools for implementing solutions to the identified problems. \cite{b19}. This suggests that technology bridges the gap between software development and the implementation of quality software solutions. CS focuses on implementing mathematical theorems/algorithms designing solutions to a problem, however SE focuses on the use of technology and computers as a problem-solving tool which thus suggests that SE has made distinctive features from CS.\\\\
In academia, programming principles and basic CS theory are often required in both CS and SE disciplines, beyond these essential aspects, they differ in what they emphasize. The core of the CS discipline is kept small and allows individuals to learn about various more advanced areas such as systems, networking, database, artificial intelligence, theory, machine learning, etc and create new knowledge of computers \cite{b23}. Whereas SE discipline, on the other hand, is often focused on rigorous techniques for designing and producing software applications that consistently perform computing actions they are programmed to execute.\\\\
CS has not been successful in training professionals with expertise in developing sophisticated software systems that meet the technical skill requirements beyond programming foundations. Information Technology (IT) discipline focuses on the study of solutions and needs that connect people, information, and the technology of the time \cite{b33}. Whereas SE focuses on large-scale software development processes, especially in areas where safety and security are crucial. IT professionals leverage the software application tools developed by software engineers in providing technological solutions that meet users' needs.\\\\
SE is an evolving engineering discipline and covers a wide application domain like problem modeling and analysis, software design, software verification and validation, software quality, software process, software management, and many more \cite{b27}. Today SE is characterized by social, cultural and human- centric issues as it provides effective team procedures as individuals need to deal with different people at various stages of software development \cite{b5}. According to Fernandez and Passoth, those issues include application domain-specific questions (e.g., on domain-specific terminologies, concepts, and procedures), ethical questions (e.g., moral assessments in the context of safety-critical situations), juridical questions (e.g., on data privacy or regulations of algorithms and their environment respectively), psychological questions (e.g., on improvements of team communications or working environments), or social and political questions (e.g., on societal impacts of software-driven technologies, the concerns of heterogeneous actors, or accountability issues). Those questions may have suggested new areas of research in SE, which focus not only on the technical elements of software development but also human needs. In addition, software systems are essential to enable world-class healthcare, commerce, education, energy generation, and many more, they have collectively weaved themselves into humankind’s daily life. There is no doubt that as technologies in diverse areas are increasingly controlled by software with the constant emergence of the push for innovation in technologies, the importance of software engineering is sure to continue growing\cite{b5}.

\section{Future Directions}
As the discipline keeps making great strides in technological development and advancement, the SE body of knowledge keeps introducing and enforcing ISO industry standards to gauge the way software is developed to meet ISO standards. Other research interest areas have come out such as Artificial Intelligence (AI), Machine Learning, and Robotics technologies, and are making a profound impact on how systems are designed and how the machine thinks.\\\\
In the book “The Information: A History A Theory A flood” Gleick (2011) highlighted the origin of IT, most researchers today have taken advantage of Shannon and Turing development of the Turing machine to step up more research into the area of cryptocurrencies, specifically blockchains and added interest in the advancement of machine learning. Also, Claude Shannon and Norbert Wiener theories form the basis of the field of AI, most organizations build systems that can test the relationship between control and communication in a self-regulating human-like machine that makes use of AI and feedback control systems by using information from their environment that they usually pick up from a sensor. In today's AI world abounds around us and is a tool for the advancement of knowledge and a means to do work with more efficiency and ease. Apple’s Siri, Amazon’s Alexa, Tesla’s self-driving cars are all innovations that today are possible based on the fundamentals in cybernetics built by Wiener and research is still evolving to advance it. There has also been an interest in building systems with no entropy. The fundamental concept of reducing information loss is important in today’s machine learning and AI. A command given to a robot must be understood appropriately (no entropy in communication) to produce the desired result. For example, Tesla’s self-driving cars process information about road conditions, surround car speeds, etc., and any loss of information (entropy) would lead to severe harm when the car is given a command to drive. Thus, Shannon’s concept of entropy in information and the drive to reduce it continues to thrive in IT all around us and to the future\cite{b32}.

\end{document}